\documentclass[twocolumn]{svjour3}         % twocolumn
\smartqed  % flush right qed marks, e.g. at end of proof
\usepackage{graphicx}
\begin{document}

\title{Using Fast-Switching Data to Characterize Atmospheric
Phase Fluctuations at the Submillimeter Array%\thanks{Grants or other notes
%about the article that should go on the front page should be
%placed here. General acknowledgments should be placed at the end of the article.}
%special issue ALMA (Dr. Bachiller).}
}
%\subtitle{Do you have a subtitle?\\ If so, write it here}

\titlerunning{Fast-switching at the SMA}        % if too long for running head

\author{Dharam Vir Lal         \and
        Satoki Matsushita    \and
        Jeremy Lim %etc.
}

\authorrunning{Lal et~al.} % if too long for running head

\institute{Dharam Vir Lal \at
              Institute for Astronomy and Astrophysics,
              Academia Sinica, P.O.~Box 23-141, Taipei 10617, Taiwan, (R.O.C.) \\
              Tel.: +886-2-3365-2200\\
              Fax: +886-2-2367-7849\\
              \email{dharam@asiaa.sinica.edu.tw}            \\
             \emph{Present email address:} dharam@mpifr-bonn.mpg.de  %  if needed
%          \and
%          S. Author \at
%             second address
}

\date{Received: date / Accepted: date}
% The correct dates will be entered by the editor

\maketitle

\begin{abstract}
For the submillimeter band observations, we have been
routinely adopting the calibration cycle time of 20--30
minutes, which is the same as any typical centimeter and millimeter
band observations.
This cycle time, largely corrects only the instrumental phase
fluctuations and there exists residual phase fluctuations, which are
attributed to temporal and spatial atmospheric phase fluctuations.
Hence, the classical calibration cycle needs closer attention for
any future submillimeter band observations.  We have
therefore obtained fast-switching test data, cycling between three
nearby calibrators, using the submillimeter array (SMA) with a cycle
time of $\sim$90 sec, in order to understand and optimize the
calibration cycle suitably, thereby to achieve the projected
sensitivity, angular resolution and dynamic range for the
SMA.  Here, we present the preliminary results from this study.
%Insert your abstract here. Include keywords, PACS and mathematical
%subject classification numbers as needed.
% \PACS{PACS code1 \and PACS code2 \and more}
% \subclass{MSC code1 \and MSC code2 \and more}
\end{abstract}

\section{Introduction}
\label{intro}
The Submillimeter Array (SMA, Ho et al. 2004) is the world's first
dedicated submillimeter band interferometer, located at the Mauna Kea,
Hawaii, $\sim$4100~m altitude, and consists of eight 6~m antennas.
It covers the frequency range of 180--900~GHz with a 2~GHz bandwidth
in both upper and lower side-bands.

The water vapor in our atmosphere is spatially inhomogeneous and highly
time-variable, resulting in atmospheric electrical path fluctuations that
alter the phase of propagating electromagnetic radiations (Battat et al. 2004).
At submillimeter bands, the effect of atmosphere fluctuations is
particularly severe, and limits the resolution, and the coherence
of interferometric arrays.

\section{Our Goal}
\label{sec:goal}
Fast switching entails switching between the target source and
a bright calibrator on timescales shorter than the baseline
crossing time of the water vapor clumps by the wind.
At this moment, we are adopting the calibration cycle time of
20--30 minutes, which is the same as the millimeter observations.
However, we are not sure if this is the right way for the submillimeter
observations.  The phase fluctuation is larger for the submillimeter
wave than the millimeter wave; therefore, the classical calibration cycle may
be too long for the submillimeter observations.  To determine the
optimized calibration cycle, we need to understand the characteristics
of the atmosphere ({\it e.g.}, structure function).
This characteristic information would also allow us to determine, whether we
need to do fast switching, what is the optimized integration time for
each data point (it currently being 30--60 sec), etc.  This information,
in future, will allow us to obtain data of highest possible quality.

\begin{table*}
\caption{Log of SMA test observations.  For each of the two experiments,
weather, available antennas, observing bands, the calibrator quasars observed,
and the range of baseline lengths available are listed.}
\begin{tabular}{l|l|ll}
\hline
& & & \\
& Experiment I & & Experiment II \\
& & & \\
\hline
& & & \\
Observe date & 05 Sep 2004 & & 02 Sep 2004 \\
%SMA test log entry & 7909 &  & 7885 \\
Weather & $\tau_{\rm 225~GHz}$ = 0.10--0.14 & & $\tau_{\rm 225~GHz}$ = 0.08--0.13 \\
Antennas  &  1, 2, 3, 4, 5, 6, 7, 8 & & 1, 2, 3, 4, 7, 8 \\
Receiving Bands & 230 GHz & & 345 GHz \\
Loop between quasars & 3C279, 1244$-$255, 1334$-$127 & & 3C454.3, 2230$+$114, 2145$+$067 \\
Calibration quasar & 3C279 & & 2230$+$114 \\
Flux densities & \multicolumn{1}{r|}{3C279 :  $\sim$8 Jy} & & \multicolumn{1}{r}{2230$+$114 : $\sim$5 Jy} \\
 & \multicolumn{1}{r|}{1244$-$255 : $\sim$1 Jy}  & & \multicolumn{1}{r}{3C454.3 : $\sim$4 Jy}  \\
 & \multicolumn{1}{r|}{1334$-$127 : $\sim$4 Jy}  & & \multicolumn{1}{r}{2145$+$067 : $\sim$2 Jy}  \\
Relative separations between quasars & $\sim$12--20~deg & & $\sim$06--18~deg \\
Baseline lengths, shortest/longest & $\sim$20/170~k$\lambda$ & & $\sim$30/260~k$\lambda$ \\
& & & \\
\hline
\end{tabular}
\end{table*}

\section{Proposed Experiments}
\label{sec:expts}

We therefore conducted following two experiments as test observations
using the SMA in the standard spectral line mode.
We chose three calibrators which are close-by (relative separation
between them being $\sim$6--20~deg) and made a loop cycle of
$\sim$90~sec with an integration time of $\sim$15~sec for each source.
Table~1 gives the details of the observations.  The continuum data
from the full data set were extracted for our analysis below.
For both experiments, we adopted similar observing strategy and
the data was analysed using Classic `AIPS'.
Since, the relative flux density would suffice our requirements,
we did not set the flux density scales for the observed sources.
We performed our analysis of the fast switching technique in the 
image-plane, and the imaging was performed using AIPS task IMAGR
and the `uniform' weighting function was used.

\subsection{Experiment I}
\label{sec:I2}

\begin{figure}
\begin{center}
  \includegraphics[width=8.5cm]{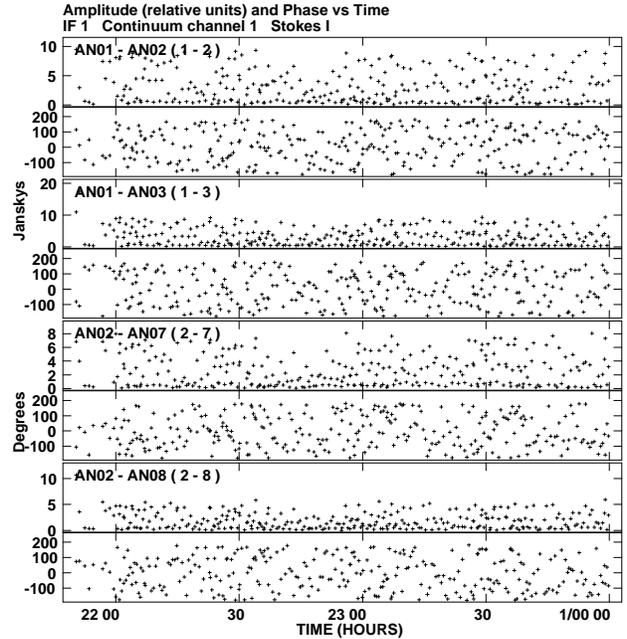}
\end{center}
\caption{Figure showing the stability of phase (in degrees) and amplitude (in Jy)
for 1244$-$255 and 1334$-$127 sources at 230~GHz for the duration of observation
(Experiment~I).}
\label{fig:I1}
\end{figure}

\begin{figure}
\begin{center}
  \includegraphics[width=8.5cm]{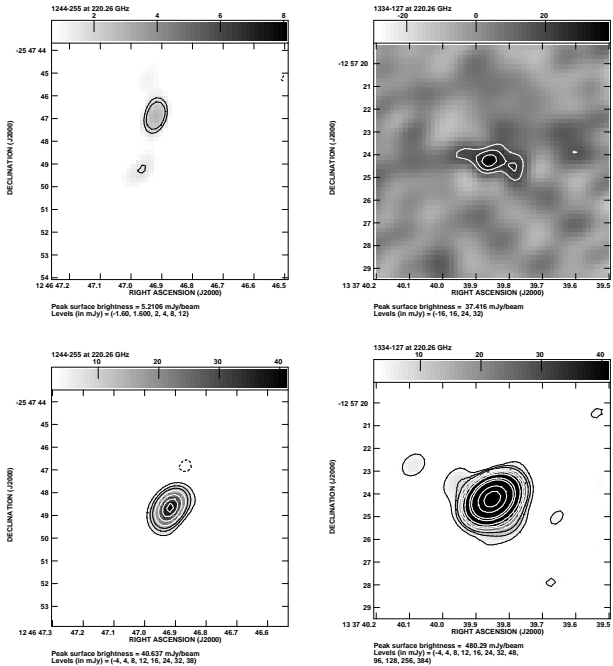}
\end{center}
\caption{Figure showing images of 1244$-$255 (left panels)
and 1334$-$127 (right panels) at 230~GHz (Experiment~I).
The upper images were reduced under the fastest switching
cycle of 90 sec and CLEANed; whereas the lower images
are the CLEAN images from the phase-only self-calibrated visibility data.
The barely detected 1244$-$255 (upper left panel) source is amplified
in the phase-only self-calibrated image (lower left panel),
implying that by using the methodology of self-calibration,
we could extract most of the flux density.}
\label{fig:I2}
\end{figure}

As a first exercise,
we observed sources 3C279, 1244$-$255 and 1334$-$127 at 230~GHz.
Typical separation between any two sources is $\sim$12--20~deg.
We chose 3C279, the brightest source, as the calibration quasar and
used it to map 1244$-$255 and 1334$-$127 at the fastest switching cycle.

The  data set suffered from severe atmospheric phase errors and
had extremely bad phase stability (see Fig.~1).
In spite of having a fastest switching cycle, namely 90~sec cycle,
the two target sources, 1244$-$255 and 1334$-$127
were barely detected at 3.5 and 2.5 sigma levels, respectively.
On the other hand, on self-calibrating (phase only) the visibility data with
its own map  and mapping the same, we could extract most of the flux density (Fig.~2).

To conclude, in this severe weather condition, fast switching cycle
of $\sim$90~sec does not work.  Instead, the observations possibly
require further shorter switching cycle.
Furthermore, since, this fastest switching cycle did not provide fruitful
results and the phase stability being bad, we did not attempt to increase
the switching cycle time.

\subsection{Experiment II}
\label{sec:II2}
Here, we observed sources 3C454.3, 2230$+$114 and 2145$+$067 at 345~GHz.
Typical separation between any two sources is $\sim$06--18~deg.
Once again, here, all three observed sources are of different flux densities,
we chose 2230$+$114, the strongest source as the calibration quasar and
used it to map 3C454.3 and 2145$+$067 at the fastest switching cycle.
In order to increase the switching cycle time, the interleaved observations
of calibrator source 2230$+$114 were FLAGged, thereby we obtained
switching cycle in multiples of 1, 2, 3, 4, 5, 6, 7, 8, 9, 10, 11, 12,
13, 14 and 15 times $\sim$90~sec.

As compared to the earlier 230 GHz data, this  data set did not suffer
from severe atmospheric phase errors (see Fig.~3).  Instead, it had a
better phase stability.  It made very little difference having a fastest
switching cycle or the default switching cycle, and in either case
the two target sources were detected at more that 10 times the noise levels.
Furthermore, on self-calibrating (phase only) the visibility data
with its own map and mapping the same, we could extract most of the
flux density and the achieved dynamic range was more than 20 in the two maps
(Fig.~4).

To conclude, in this good weather condition, fast switching cycle of
$\sim$90~sec and $\sim$22 min does not have significant difference,
but the self-calibration (phase-only) improves significantly.  
Therefore, to meet the signal-to-noise levels of maps obtained {\it via.},
self-calibration, a much shorted switching cycle is needed, and
hence to improve the map quality.

\begin{figure}
\begin{center}
  \includegraphics[width=8.5cm]{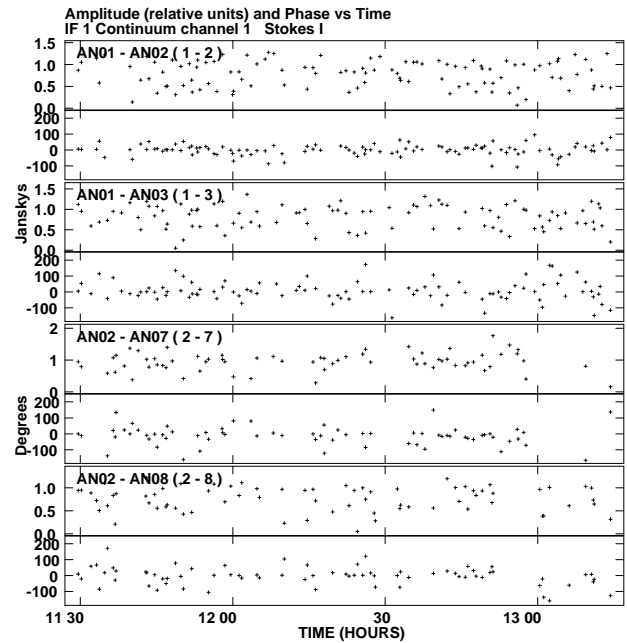}
\end{center}
\caption{Figure showing the stability of phase (in degrees) and
amplitude (in Jy) for 3C454.3 and 2145$+$067 sources at 345~GHz for
the duration of observation (Experiment~II).
This weather condition is good for science observations.}
\label{fig:II1}
\end{figure}

\begin{figure}
\begin{center}
  \includegraphics[width=8.5cm]{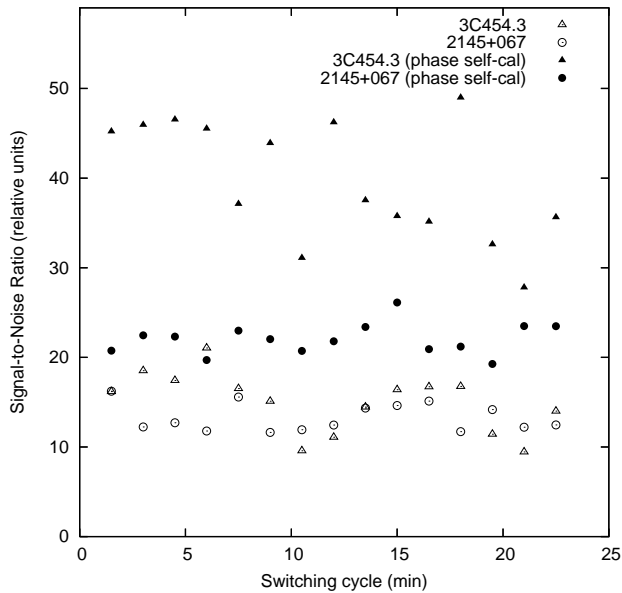}
\end{center}
\caption{Signal-to-noise ratio based on series of images of
3C454.3 and 2145$+$067 sources at 345~GHz as a function of switching cycle
(Experiment~II).
The left-most data point has ``1~times $\sim$90~sec" as the switching cycle,
subsequently increasing in multiples of
1, 2, 3, 4, 5, 6, 7, 8, 9, 10, 11, 12, 13, 14, and 15.
The lower two sets of symbols (open circles and open triangles) are from the
CLEAN images, whereas the upper two sets of symbols (filled circles and
filled triangles) are from the phase-only self-calibrated CLEAN images.}
\label{fig:II2}
\end{figure}

\section{Discussion}
\label{sec:discuss}
Phase variations at the SMA, {\it i.e.} at observing frequencies
$\le$ 400~GHz, are larger than centimeter and millimeter band observation,
and are caused predominantly by temporal changes in the water vapor content.
The implied changes in index of refraction are non-dispersive,
in particular at millimeter bands ({\it e.g.} Sutton \& Hueckstaedt 1996,
Pardo et al. 2001), and hence phase variations increases linearly with
frequency.
 
One of the methods to compensate these phase
fluctuations is the fast switching method.
%
%Furthermore, the opacity at centimeter wavelengths is mainly due to the
%dry atmosphere (molecular oxygen), which does not vary substantially either
%on diurnal timescales, while the liquid water content (clouds) contributes
%opacity (due to scattering), but does not alter phases substantially.
%
The technique of phase calibration by fast switching has been successfully
demonstrated with the Very Large Array (VLA) at 22~GHz with the minimum
cycle time being 80~sec in the standard VLA mode (Holdaway et~al. 1995)
and subsequently improved to 40~sec at 22 and 43~GHz (Carilli et~al. 1996).
%
%One of the key question to address when implementing fast switching as
%standard observing mode is:  are there enough calibrators in the sky
%in order to take advantage of a fastest possible switching time?
%
Therefore, the phase calibration through fast switching is important,
in particular, when the dynamic range limitation is set by the phase errors
and/or to obtain images of faint sources with diffraction limited resolution
on arbitrarily long baselines.

However, the fast switching technique in our two experiments did
not work as planned.
One possible reason could be that the baseline lengths used in
the two experiments are small (in the two experiments conducted,
the longest baselines were $\sim$170 k$\lambda$ at 230 GHz and
$\sim$260 k$\lambda$ at 345 GHz)
and hence, the effect of atmosphere is not completely checked.
Or in other words, it seems that the diffraction limited images
of the quasars are even more smaller and instead,
the obtained spatial resolution is limited by tropospheric `seeing'
(Morita et~al. 2000, Carilli \& Holdaway, 1999, 1997,
Carilli et~al. 1996, Holdaway et~al. 1995).
It is also possible that during observations either too bad or
too good weather conditions is unlikely to provide any improved results.

%For the sake of completeness, we mention here that we tried
%to keep all possible longest baselines during our fast switching experiments,
%in order to allow for diffraction limited imaging on the longest baselines.
%Otherwise, the spatial resolution will be limited by tropospheric `seeing'
%(Carilli et~al. 1996, Holdaway et~al. 1995).
%In our observations, the experiments were conducted with longest baselines,
%which were xy k$\lambda$ at 230 GHz and yz k$\lambda$ at 345 GHz.

In addition, in these and future observations,
we should also consider the effect of natural weighting
%alongwith the robustness parameter
during imaging for the fast switching technique.  Since the weather condition
during the experiment~II was good and$/$or the phase stability being good, 
hence the improvement in the image-plane will not be as dramatic as
it may be in the case of experiment~I.

\section{Conclusions}
\label{sec:conclude}
The data presented above attempts to provide a measure of the atmospheric
phase fluctuations between antennas and can help recover suitable results
depending on intrinsic atmospheric phase fluctuations.

\begin{enumerate}
\item[{$\bullet$}] Based on these two experiments, both, very bad
weather conditions and good weather conditions,
we have demonstrated that there is no
difference between $\sim$90 sec and $\sim$22~min calibration switching cycle.
However, self-calibration works better in both cases, suggesting that a
further shorter calibration switching cycle is needed in both weather
conditions.
\end{enumerate}

We conclude that the self-calibration can often be performed
(typically with a 40~sec averaging time on sources stronger
than 100~mJy at 43~GHz (Carilli et~al. 1996) and $\sim$5~sec
averaging time on sources observed here).  Furthermore,
there is definitely a need to undertake an additional experiment in order
to bridge the gap between these two extreme experiments.

%% For one-column wide figures use
%\begin{figure}
%% Use the relevant command to insert your figure file.
%% For example, with the graphicx package use
%  \includegraphics{example.eps}
%% figure caption is below the figure
%\caption{Please write your figure caption here}
%\label{fig:1}       % Give a unique label
%\end{figure}
%%
%% For two-column wide figures use
%\begin{figure*}
%% Use the relevant command to insert your figure file.
%% For example, with the graphicx package use
%  \includegraphics[width=0.75\textwidth]{example.eps}
%% figure caption is below the figure
%\caption{Please write your figure caption here}
%\label{fig:2}       % Give a unique label
%\end{figure*}
%%
%% For tables use
%\begin{table}
%% table caption is above the table
%\caption{Please write your table caption here}
%\label{tab:1}       % Give a unique label
%% For LaTeX tables use
%\begin{tabular}{lll}
%\hline\noalign{\smallskip}
%first & second & third  \\
%\noalign{\smallskip}\hline\noalign{\smallskip}
%number & number & number \\
%number & number & number \\
%\noalign{\smallskip}\hline
%\end{tabular}
%\end{table}
%

\begin{acknowledgements}
The Submillimeter Array is a joint project between the Smithsonian
Astrophysical Observatory and the Academia Sinica Institute of Astronomy
and Astrophysics and is funded by the Smithsonian Institution and
the Academia Sinica.
DVL thanks PTP~Ho for discussions and several useful comments.
\end{acknowledgements}

% BibTeX users please use one of
%\bibliographystyle{spbasic}      % basic style, author-year citations
%\bibliographystyle{spmpsci}      % mathematics and physical sciences
%\bibliographystyle{spphys}       % APS-like style for physics
%\bibliography{}   % name your BibTeX data base

% Non-BibTeX users please use

\end{document}